# Landau Quantization and Fermi Velocity Renormalization in Twisted Graphene Bilayers

Long-Jing Yin[1,2,§], Jia-Bin Qiao[1,2,§], Wei-Jie Zuo[1,2,§], Wei Yan[1,2], Rui Xu[1,2], Rui-Fen Dou[1], Jia-Cai Nie[1], and Lin He[1,2,*]

[1] Department of Physics, Beijing Normal University, Beijing, 100875, People's Republic of China
[2] The Center of Advanced Quantum Studies, Beijing Normal University, Beijing, 100875, People's Republic of China
[§]These authors contributed equally to this work.
* Email: helin@bnu.edu.cn

**Currently there is a lively discussion concerning Fermi velocity renormalization in twisted bilayers and several contradicted experimental results are reported. Here we study electronic structures of the twisted bilayers by scanning tunneling microscopy (STM) and spectroscopy (STS). The interlayer coupling strengths between the adjacent bilayers are measured according to energy separations of two pronounced low-energy van Hove singularities (VHSs) in the STS spectra. We demonstrate that there is a large range of values for the interlayer interaction in different twisted bilayers. Below the VHSs, the observed Landau quantization in the twisted bilayers is identical to that of massless Dirac fermions in graphene monolayer, which allows us to measure the Fermi velocity directly. Our result indicates that the Fermi velocity of the twisted bilayers depends remarkably on both the twisted angles and the interlayer coupling strengths. This removes the discrepancy about the Fermi velocity renormalization in the twisted bilayers and provides a consistent interpretation of all current data.**

The electronic properties of graphene bilayers depend sensitively on their stacking orders [1-8]. The two common stacking configurations of bilayers, *i.e.*, the AB (Bernal) and AA-stacked bilayers, exhibit very different electronic properties even though there is only subtle distinction between their stacking orders. Very recently, the introduction of a stacking misorientation has further expanded the allotropes of the graphene bilayers dramatically and, more importantly, these resultants have been demonstrated to show strong twist-dependent electronic spectra and properties [2,4-20]. As an example, the theoretical calculations predicted pronounced Fermi velocity renormalization in the twisted bilayers with small twisted angles [5-8]. The strongest experimental evidence for the Fermi velocity renormalization is the observed Fermi velocity $v_F = 0.87 \times 10^6$ m/s, basing on Landau-level spectroscopy, in the twisted bilayer with a twisted angle $\theta \sim 3.5°$ on a graphite surface [15]. However, several contradicted experimental results concerning the Fermi velocity in the twisted bilayers are reported [9,21,22]. The reducing of the Fermi velocity on twisted angles and experimental conditions is strongly debated.

In this Letter, we study electronic properties of the twisted graphene bilayers by scanning tunneling microscopy (STM) and spectroscopy (STS). The interlayer interaction strengths $t_\theta$ of the twisted graphene bilayers are measured according to energy separations of the low-energy van Hove singularities (VHSs), which originate from two saddle points in the band structures, in the tunneling spectra. The Fermi velocities $v_F$ of the twisted bilayers are measured directly according to the Landau quantization in high magnetic fields. Our result demonstrates that the Fermi velocities not only depend on the twisted angles, but also on the interlayer coupling strengths remarkably. This could provide a consistent interpretation of all current data.

The STM system was an ultrahigh vacuum scanning probe microscope (USM-1500) from UNISOKU (see Supplementary Information [23] for experimental method). The surface few-layer of highly oriented pyrolytic graphite (HOPG) usually decouples from the bulk [2,19,24-27] and the twisted graphene bilayers with various rotation angles can be easily observed on HOPG surface [19]. Therefore, the surface of HOPG provides a natural ideal platform to probe the electronic spectra of the twisted bilayers.

Figure 1(a)-(c) shows typical STM images of several twisted graphene bilayers on a HOPG surface. The twisted angle $\theta$ between neighboring graphene layers can be obtained from the period of the Moiré patterns $D$, using $D = a/[2\sin(\theta/2)]$ with the graphene lattice constant $a \sim 0.246$ nm [2,9]. The twisted angle can also be measured according to the fast Fourier transforms (FFT) of the STM images, as shown in Fig. 1(d). In the FFT image, the rotation angle $\varphi$ between the atomic lattice and the Moiré super-lattice has a relation with twisted angle $\theta$ as $\varphi = 30° - (\theta/2)$. In our experiment, the obtained twisted angles of the adjacent graphene bilayers by the two methods are consistent.

To study the effects of the twisted angle and interlayer coupling strength on the electronic spectra, we measure tunneling spectra of different twisted graphene bilayers, as shown in Fig. 2(a) (see Supplementary Information [23] for discussion). Fig. 2(b) shows representative spatial resolved spectra of a twisted bilayer, which confirms the reproducibility of the tunneling spectra. For the bilayers with $\theta = 6.0°$ and $\theta = 8.2°$, we cannot detect the VHSs in the tunneling spectra in an $\approx 2$ eV wide region. The absence of the VHSs is attributed to decoupling behavior of twisted bilayers with large rotation angles, *i.e.*, $t_\theta \approx 0$ meV [15,20]. For the bilayers with $\theta \leq 3.8°$, two peaks in the spectra, which are attributed to the two low-energy VHSs in twisted bilayers [2,9,14,15], are observed.

Theoretically, the energy difference of the VHSs, $\Delta E_{VHS}$, can be roughly estimated according to $\Delta E_{VHS} \approx \hbar v_F \Delta K - 2t_\theta$. Here, $|\Delta \boldsymbol{K}| = 2|\boldsymbol{K}|\sin(\theta/2)$ is the relative shift of the Dirac points of the adjacent bilayers with $K$ the reciprocal-lattice vector. For a constant $t_\theta$, therefore, it is expected that $\Delta E_{VHS}$ should decrease monotonously with $\theta$. However, the observed $\Delta E_{VHS}$ of the twisted bilayers, as shown in Fig. 2(a), does not decrease monotonously with the rotation angles. In Fig. 2(d), we summarize the measured $\Delta E_{VHS}$ as a function of the twisted angles. Besides the twisted angle, the interlayer coupling strength also plays an important role in determining the value of $\Delta E_{VHS}$. Such a non-monotonic dependence between $\Delta E_{VHS}$ and $\theta$ should be attributed to variations of the interlayer coupling in different bilayers. With identical twisted angle, the energy difference of the saddle points of the twisted bilayer with a weaker coupling is much larger than that of the bilayer with a stronger interlayer coupling, as shown in Fig. 2(c) (see Supplementary Information [23] for details of calculation). The interlayer hopping $t_\theta$ of different bilayers can be estimated by comparing the measured $\Delta E_{VHS}$ with the theoretical band structures of twisted bilayers. In Fig. 2(d), we summarize the obtained $t_\theta$ in our experiment as a function of the twisted angles. Obviously, the $t_\theta$ is not a constant and there is a large range of values for the interlayer interaction in different twisted bilayers. The interlayer hopping of adjacent bilayer depends, in the simplest approximation, exponentially on the interlayer distance. In the twisted bilayers, the stacking fault, tilt grain boundary, and roughness of substrate may affect the interlayer distance and stabilize it at various equilibrium distances, resulting in the large variations of the interlayer interaction.

To further explore the electronic properties of the twisted bilayers, we perform STS measurements of these samples in different magnetic fields. Our experimental results indicate that both the twisted angles and the interlayer coupling strengths affect the Landau quantization in the twisted bilayers. Figure 3 displays several representative experimental data and analysis of the Landau quantization (see Fig. S1-S5 of Supplementary Information [23] for more experimental data and analysis). In the twisted bilayers, it is expected to generate two series of Landau level (LL) sequences below the VHSs in high magnetic fields. Each of the LL sequence depends on the square-root of both level index $n$ and magnetic field $B$ [15,25,27]:

$$E_n = E_D + \text{sgn}(n)\sqrt{2e\hbar v_F^2 |n| B}, \qquad n = 0, \pm 1, \pm 2, \ldots \qquad (1)$$

Here $E_D$ is the energy of Dirac point and $e$ is the electron charge. The sequence of LLs shown in Fig. 3(a)-3(c) is unique to massless Dirac fermions and is expected to be observed in the twisted bilayers [15]. In the experiment, we only observe one LLs sequence since that the quasiparticles of the topmost graphene layer are mainly coming from one of the two Dirac cones and the STM probes predominantly the top layer. The Fermi velocities of the twisted bilayers can be obtained directly by a reasonable linear fitting of the experimental data to Eq. (1), as shown in Fig. 3(d)-3(f).

In the literature, the theoretical calculations predicted pronounced Fermi velocity renormalization in the twisted bilayers with small twisted angles [5-8]. Figure 4(a) shows band structures of three twisted bilayers with an identical interlayer coupling strength (see Supplementary Information [23] for details of calculation). The Fermi velocity decreases with decreasing the twisted angles. The renormalization of the Fermi velocity in twisted bilayers can be described by

$$\frac{\tilde{v}_F}{v_F} = 1 - 9\left(\frac{t_\theta}{\hbar v_F \Delta K}\right)^2. \qquad (2)$$

However, a reasonable linear fitting of the experimental data of the twisted bilayer with $\theta \sim 2.8°$ to Eq. (1), as shown in Fig. 3(d), is found yielding the Fermi velocity $v_F$ = (1.03 ± 0.02) × $10^6$ m/s. Obviously, there is no significant reduction of the Fermi velocity in this sample. Here we should point out that the pronounced Fermi velocity renormalization in the TB was predicted previously basing on a strong interlayer coupling ($t_\theta \sim 110$ meV). Theoretically, when the interlayer hopping is relative weak, we still can obtain the main features of the low-energy band structure of the twisted bilayers, *i.e.*, the two displaced Dirac cones and the two VHSs, but without the significant reduction of the Fermi velocity, as shown in Fig. 3(c). We attribute the deviation between our experimental result and that reported in Ref. [15] mainly to the different interlayer hopping of the samples. In the twisted bilayer with $\theta \sim 2.8°$, the interlayer hopping $t_\theta$ is estimated to be about 50 meV by comparing the measured $\Delta E_{\text{VHS}}$ with the theoretical result. In the twisted bilayer with $\theta \sim 3.5°$, as reported in Ref. 15, $t_\theta$ is estimated to be 110 meV. Consequently, we observed a negligible renormalization of the Fermi velocity in the twisted bilayer with $\theta \sim 2.8°$, whereas the authors of Ref. 15 reported a pronounced reduction of the Fermi velocity. Such a conclusion has been further confirmed by the high magnetic fields spectra measurements of other twisted bilayers, as shown in Fig. 3. In the twisted bilayer with $\theta \sim 3.6°$, where the $t_\theta$ is as large as about 130 meV, the Fermi velocity is reduced to $v_F = (0.811 \pm 0.004) \times 10^6$ m/s, as shown in Fig. 3(e).

To clearly illustrate the effect of the interlayer coupling on the Fermi velocity of the twisted bilayers, we plot the measured Fermi velocities of the twisted bilayers with different twisted angles in Fig. 4(b) (see Fig. S1-5 of Supplementary Information [23] for more experimental data). Figure 4(b) also shows the predicted angle dependence of the renormalization for different interlayer hopping strengths, as described by Eq. (2). The interlayer hopping strengths of these twisted bilayers are obtained according to the measured $\Delta E_{VHS}$. The observed Fermi velocity does not decrease monotonously with decreasing the twisted angle, indicating that it is not only determined by the twisted angle. Carefully examining the interlayer hopping of these samples reveals that the coupling strength between the two layers plays an important role in affecting the Fermi velocity. The measured Fermi velocity agrees well with the calculated result by taking into account the interlayer hopping of the twisted bilayers, as shown in Fig. 4(b). When $t_\theta \leq 50$ meV, there is only a slight reduction of the Fermi velocity for the twisted bilayers with $\theta > 2.0°$, which may account for the negligible Fermi velocity renormalization observed in twisted bilayers previously.

In summary, we address the electronic spectra and their Landau quantization in the twisted bilayers. We demonstrate that the interlayer coupling strength affects both the VHSs and the Fermi velocity of twisted bilayers dramatically. Our result indicates that a small twist could make a graphene monolayer decouple from the substrate (the graphite here) sufficiently for electronic energy below the VHSs. This finding may have broad implications in tuning electronic properties of other two-dimensional van der Waals structures.

**Acknowledgments**

We thank Prof. A. K. Geim and Prof. E. Y. Andrei for helpful discussions. This work

was supported by the National Basic Research Program of China (Grants Nos. 2014CB920903, 2013CBA01603, 2013CB921701), the National Natural Science Foundation of China (Grant Nos. 11422430, 11374035, 11474022, 51172029, 91121012), the program for New Century Excellent Talents in University of the Ministry of Education of China (Grant No. NCET-13-0054), Beijing Higher Education Young Elite Teacher Project (Grant No. YETP0238).

**Figure Captions**

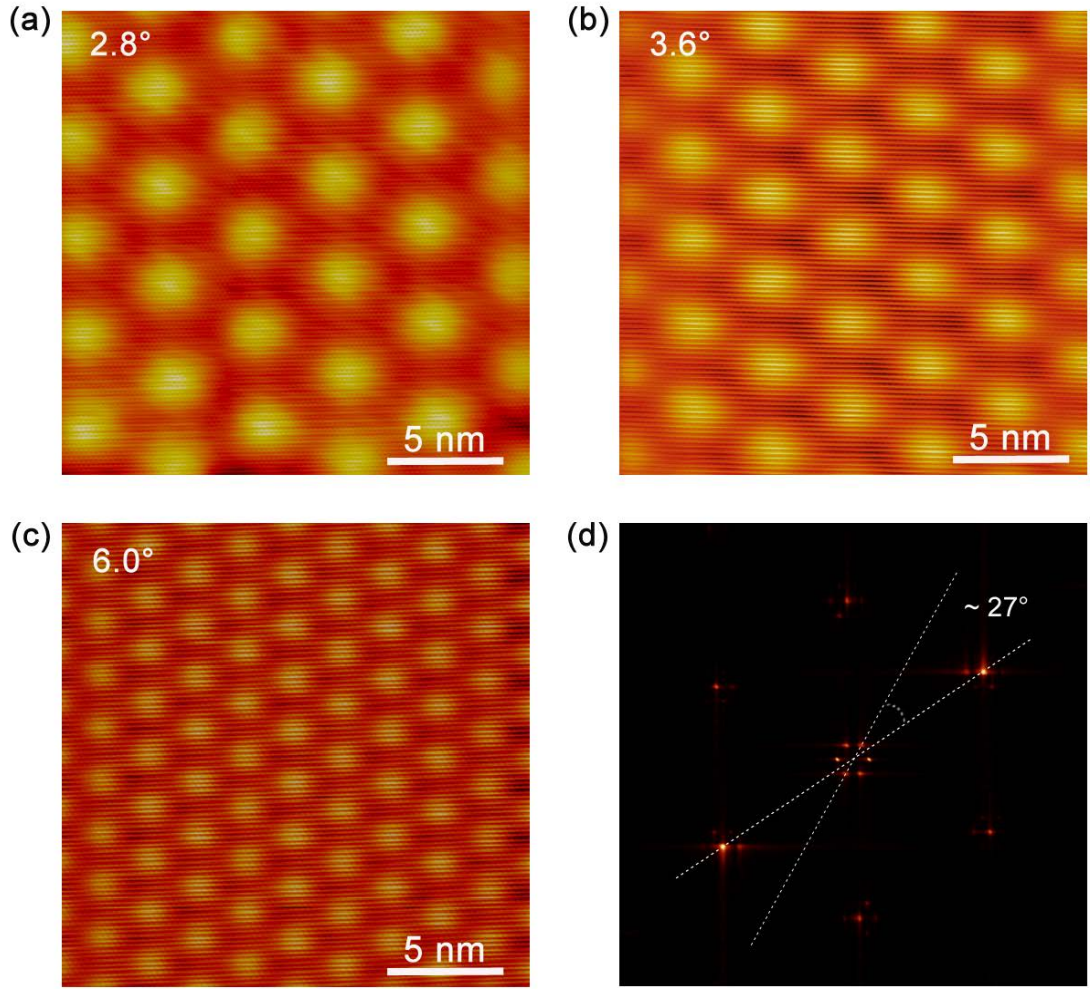

**Figure 1.** (a)-(c) 20 nm × 20 nm STM topographic images of twisted bilayers with $\theta$ = (2.8 ± 0.1)°, $D$ = 5.0 ± 0.2 nm (a); $\theta$ = (3.6 ± 0.1)°, $D$ = 3.92 ± 0.08 nm (b); and $\theta$ = (6.0 ± 0.1)°, $D$ = 2.35 ± 0.05 nm (c) on graphite surface. (d) Fast Fourier transforms image of panel (c). The outer and inner hexangular bright spots correspond to the reciprocal lattice of the graphene lattice and moiré patterns, respectively. The rotation angle $\varphi$ between the two reciprocal lattices is 27.0 ± 0.1°. The twisted angle can be estimated as $\theta = (60° - 2\varphi) = (6.0 ± 0.2)°$, which is consistent with that measured in real space.

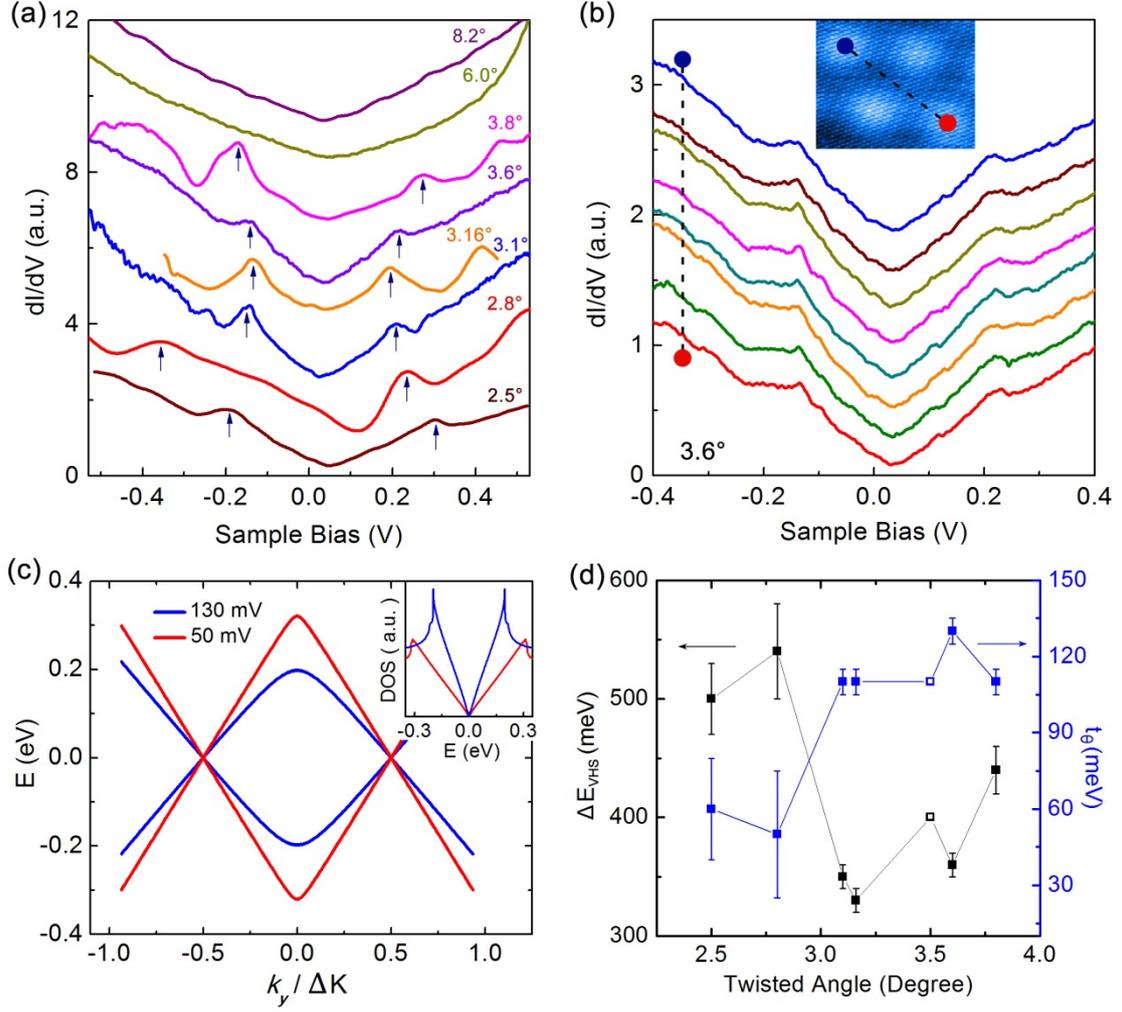

**Figure 2**. (a) STS spectra of twisted graphene bilayers with various twisted angles. From bottom to top: $\theta = (2.5 \pm 0.1)°$, $D = 5.6 \pm 0.3$ nm; $\theta = (2.8 \pm 0.1)°$, $D = 5.0 \pm 0.2$ nm; $\theta = (3.1 \pm 0.1)°$, $D = 4.5 \pm 0.2$ nm; $\theta = (3.16 \pm 0.05)°$, $D = 4.46 \pm 0.05$ nm; $\theta = (3.6 \pm 0.1)°$, $D = 3.92 \pm 0.08$ nm; $\theta = (3.8 \pm 0.1)°$, $D = 3.7 \pm 0.1$ nm; $\theta = (6.0 \pm 0.1)°$, $D = 2.35 \pm 0.05$ nm; and $\theta = (8.2 \pm 0.4)°$, $D = 1.72 \pm 0.08$ nm. (b) STS spectra of the 3.6° twisted bilayer measured in different positions along the line in the moiré pattern (see the STM image in the inset). (c) Low-energy band structures of the 3.1° twisted bilayer with interlayer hopping $t_\theta = 130$ meV and 50 meV. The inset shows the corresponding DOS of the twisted bilayer with different $t_\theta$. (d) Energy differences of the VHSs and interlayer hopping obtained in our experiment as a function of the rotation angles in twisted graphene bilayers. The empty squares are the data of Ref. 15.

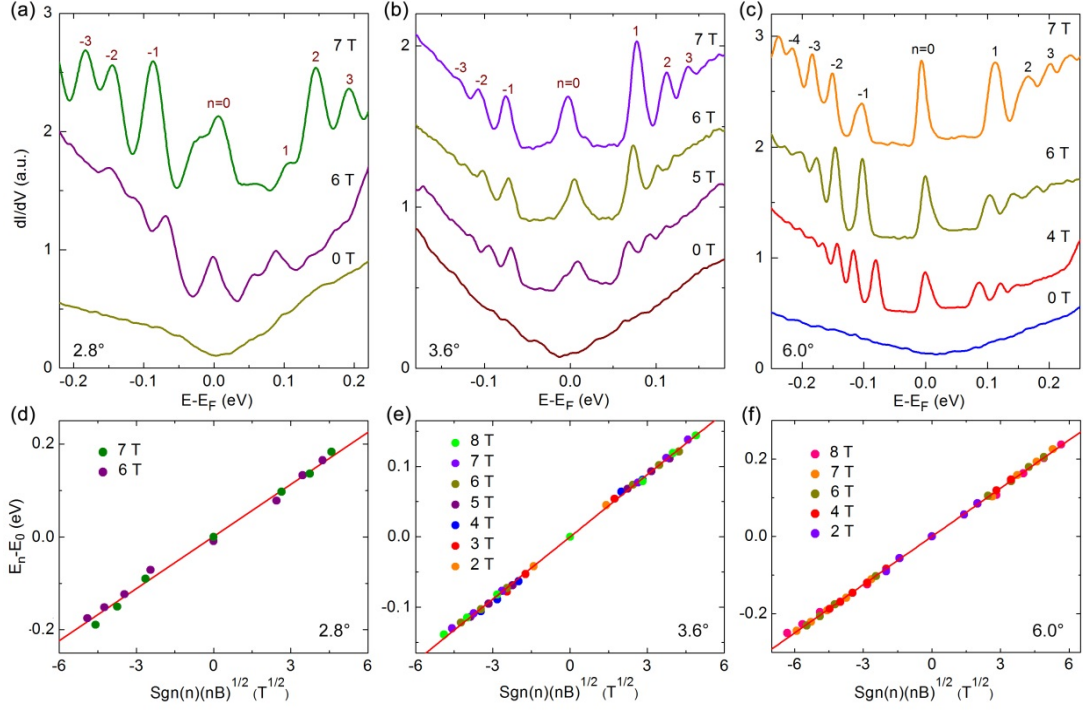

**Figure 3**. (a)-(c) STS spectra taken in various magnetic fields in twisted bilayers with $\theta = (2.8 \pm 0.1)°$ (a), $\theta = (3.6 \pm 0.1)°$ (b), and $\theta = (6.0 \pm 0.1)°$ (c). LL indices are marked. For clarity, all the spectra are offset on the Y axis, and all the curves are shifted to make the $n = 0$ LLs stay at the same bias. (d)-(f) Landau level peak energies for different magnetic fields obtained in (a)-(c) plotted against $\text{sgn}(n)(|n|B)^{1/2}$, as expected for massless Dirac fermions. The solid red lines are linear fitting of the data with the slope yielding the Fermi velocity of $v_F = (1.03 \pm 0.02) \times 10^6$ m/s (d), $v_F = (0.811 \pm 0.004) \times 10^6$ m/s (e), and $v_F = (1.140 \pm 0.003) \times 10^6$ m/s (f).

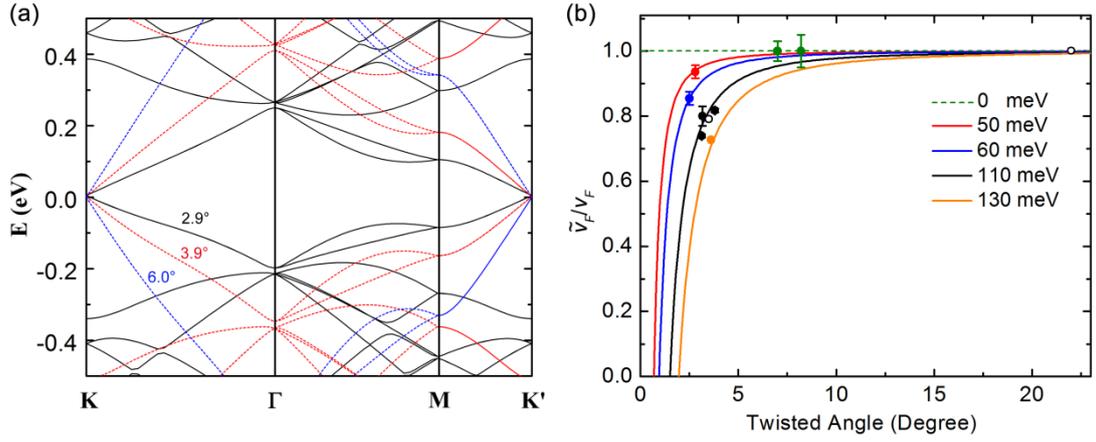

**Figure 4**. (a) Energy band dispersions for three twisted graphene bilayers with $\theta =$ 6.0°, $\theta = 3.9°$, and $\theta = 2.9°$, respectively. In the calculation, the interlayer coupling strength for the three twisted bilayers is identical. The Fermi velocity is reduced with decreasing the twisted angles. (b) The Fermi velocity as a function of the twisted angles for different interlayer interactions. The solid circles are the Fermi velocities of the twisted bilayers obtained in our experiments. The open circles are the experimental data reported in Ref. 15. The curves are the predicted Fermi velocity renormalization with different interlayer hopping strengths. Here, we use different colors to denote different interlayer hopping strengths and the Fermi velocity is normalized with respect to $v_F = 1.10 \times 10^6$ m/s.